\documentclass[epj,english,final,amsmath,amssymb]{svjour}
\usepackage{graphicx}
\input{epsf}
\begin{document}
\title{Dissipative decoherence in the Grover algorithm}

\author{O. V. Zhirov\inst{1}  and D. L. Shepelyansky\inst{2}}

\institute {Budker Institute of Nuclear Physics, 630090 Novosibirsk, Russia
\and
Laboratoire de Physique Th\'eorique,
UMR 5152 du CNRS, Univ. P. Sabatier, 31062 Toulouse Cedex 4, France}

\titlerunning{Dissipative decoherence in the Grover algorithm}
\authorrunning{O.V.Zhirov and D.L.Shepelyansky}

\date{Received:}
\abstract{Using the methods of quantum trajectories
we study effects of dissipative decoherence  on the accuracy
of the Grover quantum search algorithm.
The dependence on the number of qubits and
dissipation rate are determined and tested numerically
with up to 16 qubits. As a result,
our numerical and analytical studies give
the universal law for decay of fidelity and 
probability of searched state  which are induced by dissipative
decoherence effects. This law is in agreement with
the results obtained previously for quantum chaos algorithms.
}

\date{November 2, 2005}

\maketitle


Nowadays the quantum computing attracts a great interest
of the scientific community \cite{nielsen}.
The main reason of that is due to the fact that certain quantum
algorithms allow to perform computations much faster
than the usual classical algorithms.
The famous example is  the Shor algorithm
which performs factorization of integers on a quantum computer  
exponentially faster than any known classical
factorization algorithm \cite{shor}.   
However, at present there is no mathematical prove
about efficiency of potentially possible classical algorithms
that in this case gives certain restrictions on the comparative
efficiency of classical and quantum algorithms.
The situation is different in the case of the Grover
quantum search algorithm \cite{grover}.
Indeed, it has been proved that it is quadratically
faster than any classical algorithm 
(see e.g. \cite{nielsen} and Refs. therein).

In addition to the question of quantum algorithm efficiency
it is also important to know what is the accuracy
of a quantum algorithm in presence of realistic 
errors and imperfections. 
The accuracy can be characterized by the fidelity $f$ \cite{peres}
defined as a scalar product of the wave function of an ideal
quantum algorithm and the wave function
given by a realistic algorithm (see e.g. \cite{nielsen}).
In general it is possible to distinguish
three types (classes) of errors. The first class can be viewed
as random unitary errors in rotational angles of
quantum gates. This is the mostly studied case
which has been also analyzed for various
quantum algorithms with the help of numerical simulations
with up to 28 qubits (see e.g. \cite{zoller,paz,cat,terraneo,bettelli,frahm}).
It has been shown that in such a case the fidelity
decays exponentially with the number of quantum gates $n_g$ and
with a rate $\gamma$ which is proportional to 
a mean square of fluctuations in gate rotations.
The second class of errors is related to static imperfections
(static in time). They are produced by static residual 
couplings between qubits and static energy shifts of individual qubits
which may generate many-body quantum chaos in a quantum
computer hardware \cite{georgeot00}.
This second type of errors, e.g. static imperfections, 
gives a more rapid decay of
fidelity as it has been shown in \cite{benenti,frahm}.
In the case of quantum algorithms for a complex dynamic
these imperfections lead to
the fidelity decrease described by a universal decay law
given by the random matrix theory \cite{frahm}. 
The two former classes are related to unitary errors.
However, there is also the third class which corresponds to
the case of nonunitary errors typical to 
the case of dissipative decoherence.
This type of errors has been studied recently for the
quantum baker map \cite{carlo} and the quantum sawtooth
map \cite{lee}. It has been shown that the exponential
decay rate of fidelity is proportional to the number of qubits.
The dissipative decoherence is treated in the frame of
Lindblad equation for the density matrix \cite{lindblad}.
A relatively large number of qubits can be reached
by using the methods of quantum trajectories developed
in \cite{qt1,qt2,qt3,qt4,qtr1,qtr2,qtr3}.

The quantum algorithms studied in 
\cite{cat,terraneo,bettelli,frahm,benenti,carlo,lee}
describe quantum and classical evolution of dynamical systems.
However, it is also important to analyze 
the accuracy
of realistic quantum computations for  
more standard algorithms, e.g. for the Grover algorithm.
The effects of unitary errors of the first and  second
classes have been studied in \cite{song} and \cite{zhirov}
respectively. It has been shown that the accuracy
of computation is qualitatively different 
in case of random errors in gates rotations \cite{song}
and in the case of static imperfections \cite{zhirov}.
Thus it is important to analyze the effects of dissipative
decoherence in the Grover algorithm
to have a complete picture for this well known algorithm.
For that aim we use the approach developed in \cite{carlo,lee}.

To study the effects of dissipative decoherence in the Grover
algorithm we use the same notations as in \cite{zhirov}.
Thus, the computational basis of a quantum register with $N=2^{n_q}$ states
($\lbrace\vert x\rangle \rbrace $, $x=0,\ldots,N-1$)
is used for the algorithm itself.
According to \cite{nielsen,grover}, the initial state 
$\vert \psi_0 \rangle=\sum_{x=0}^{N-1}\vert x \rangle/\sqrt{N}$,
is transfered to the state
\begin{equation}
\label{eq1}
  \begin{array}{ll}
       \vert \psi(t) \rangle=\hat{G}^t\vert \psi_0 \rangle \\
           =\sin{((t+1/2) \omega_G)}\vert \tau \rangle + 
\cos{((t+1/2) \omega_G)}\vert \eta \rangle \; ,
   \end{array}
\end{equation}
after $t$ applications of the Grover operator $\hat{G}$.
Here, $\omega_G=2\arcsin(\sqrt{1/N})$ is
the Grover frequency,  $\vert \tau \rangle $ is the search state and
$\vert \eta \rangle=\sum_{x\neq\tau}^{(0\leq x<N)}\vert x \rangle/\sqrt{N-1}$.
Hence, the ideal algorithm
gives a rotation in the 2D plane $(\vert \tau \rangle,\vert \eta \rangle)$.
One iteration of the algorithm is given by the Grover operator
$\hat{G}$ and can be implemented in  
$n_g=12n_{tot}-42$ elementary gates including
one-qubit rotations, control-NOT and Toffolli gates 
as described in \cite{zhirov}. The implementation of
all these gates requires  an ancilla qubit
so that the total number of qubits is $n_{tot}=n_q+1$.

To study the effects of dissipative decoherence
on the accuracy of the Grover algorithm
we follow the approach used in \cite{carlo,lee}.
The evolution of the density operator $\rho(t)$ of open system
under weak Markovian noises is given by
the master equation with Lindblad operators $L_m~(m=1, \cdots, n_{tot})$ 
\cite{lindblad}:
\begin{equation}
\label{eq2}
\dot{\rho}=-\frac{i}{\hbar}[H_{eff} \rho -\rho H_{eff}^\dagger]
+\sum_m L_m \rho  L^\dagger_m \; ,
\end{equation}
where the Grover  Hamiltonian $H_G$ ($\hat{G}=\exp(-iH_G)$)
is related to the  effective Hamiltonian
 $H_{eff}\equiv H_G-i\hbar/2 \sum_m L^\dagger_m  L_m$
and $m$ marks the qubit number.
In this paper we assume that
the system is coupled to the environment through
an amplitude damping channel with
$L_m=\hat{a}_m\sqrt{\Gamma}$,
where $\hat{a}_m$ is the destruction operator for $m-$th qubit
and the dimensionless rate $\Gamma$ gives the decay rate for
each qubit per one quantum gate.
The rate $\Gamma$ is the
same for all qubits.

This evolution of $\rho$ can be  efficiently simulated
by averaging  over the  $M$ quantum trajectories 
(see e.g. \cite{qt1,qt2,qt3,qt4,qtr1,qtr2,qtr3})
which evolve
according to the following stochastic differential equation
for states $|\psi^\alpha\rangle~(\alpha=1, \cdots , M)$:
\begin{equation}
\label{eq3}
  \begin{array}{ll}
|d\psi^\alpha\rangle=-iH_s |\psi^\alpha\rangle dt
+\frac{1}{2}\sum_m ( \langle L^\dagger_m L_m\rangle_\psi \;\; \\
 - L_m^\dagger L_m)
|\psi^\alpha\rangle dt+
\sum_m \left (\frac{L_m}{\sqrt{\langle L_m^\dagger L_m\rangle_\psi}}-1\right )|\psi^\alpha\rangle d N_m \; ,
  \end{array}
\end{equation}
where $\langle  \rangle_\psi$ represents an expectation value
on $|\psi^\alpha\rangle$ and $dN_m$ are stochastic differential variables
defined in the same way as in \cite{carlo} (see Eq.(10) there).
The above equation can be solved  numerically
by the quantum Monte Carlo (MC) methods
by letting the state $|\psi^\alpha\rangle$
jump to one of $L_m|\psi^\alpha\rangle/|L_m|\psi^\alpha\rangle|$  states
with probability 
\newline
\noindent
$dp_m\equiv |L_m|\psi^\alpha\rangle|^2 dt$ \cite{carlo}
 or evolve to
\newline
\noindent
$(1-iH_{eff} dt/\hbar)|\psi^\alpha\rangle/\sqrt{1-\sum_m dp_m}$
with probability $1-\sum_m dp_m$.
Then, the density matrix can be approximately expressed as
\begin{equation}
\label{eq4}
\rho(t) \approx \left \langle  |\psi(t)\rangle \langle \psi(t)|\right \rangle_M
= \frac{1}{M}\sum^M_{\alpha=1}|\psi^\alpha(t)\rangle  \langle\psi^\alpha(t)|
\; ,
\end{equation}
 where $\langle \rangle_M$ represents an ensemble average over $M$ quantum
 trajectories $|\psi^\alpha(t)\rangle$.
Hence, an expectation value of an operator $O$ is given by
$ \langle O \rangle=Tr(O\rho) \approx \langle O\rangle_M$.

\begin{figure}
\epsfxsize=3.2in
\epsfysize=2.6in
\epsffile{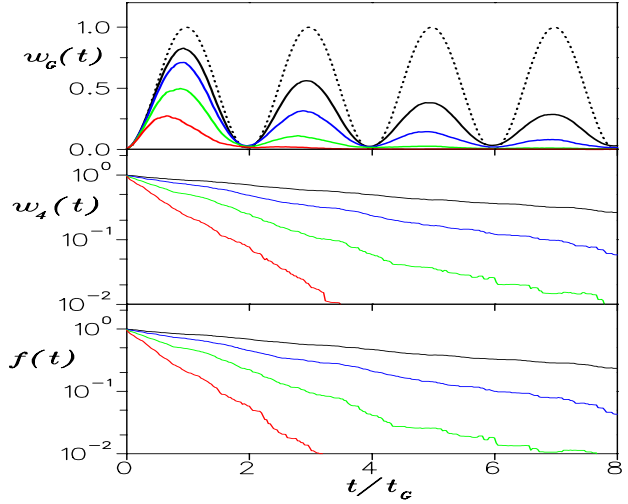}
\caption{(color online) Probability of a searched state $w_G(t)$ (top), 
    the weight of
    4-state subspace $w_4(t)$ (middle) and the fidelity $f(t)$ (bottom) 
    in the Grover algorithm
    as a function of the iteration step $t$     for $n_{tot}=12$ qubits,
    $t_G=35.5$. Dotted curve shows data for the ideal Grover algorithm, 
    and solid
    curves from top to bottom correspond to 
    one qubit decoherence rate $\Gamma=(1,2,4,8)\times 10^{-5}$.
    Number of quantum trajectories used in simulations is $M=400$.
}
\label{fig1}
\end{figure}

In presence of
dissipative decoherence
the fidelity $f$ of quantum algorithm is defined as
\begin{equation}
\label{eq5}
f(t)\equiv \langle \psi_0(t) | \rho(t) |\psi_0(t)\rangle \approx
\frac{1}{M} \sum_\alpha |\langle \psi_0(t) |\psi^\alpha_\Gamma(t)\rangle |^2
\; ,
\end{equation}
where $| \psi_0(t)\rangle $ is the wave function given by the exact
algorithm and $\rho(t)$ is the density matrix of the
quantum computer in presence of decoherence, both are taken
after $t$ map iterations. Here, $\rho(t)$ is expressed
approximately through the sum over quantum trajectories
(see also \cite{carlo}).

The effects of dissipative decoherence for the Grover algorithm with
$n_{tot}=12$ qubits are presented in Fig.~\ref{fig1}. The data
for the probability of the searched state $w_G(t)$ 
clearly show that the Grover oscillations, which have the period
$2t_G \approx \pi/\omega_G$, start to decay with the number of iterations $t$
due to decoherence. It is not so easy to extract an exponential 
decay superimposed with oscillations. Therefore, it is convenient
to study also the decay of the 4-states probability $w_4(t)$
($\vert \tau \rangle$ and  $\vert \eta \rangle$
for two states of ancilla qubit, see \cite{zhirov}).
This probability  has a pure exponential decay
with the rate $\gamma$: $w_4(t) = \exp(-\gamma t)$.
The fidelity decay has the same rate $\gamma$ (see Fig.~\ref{fig1}).

\begin{figure}
\includegraphics[width=2.1in,angle=90]{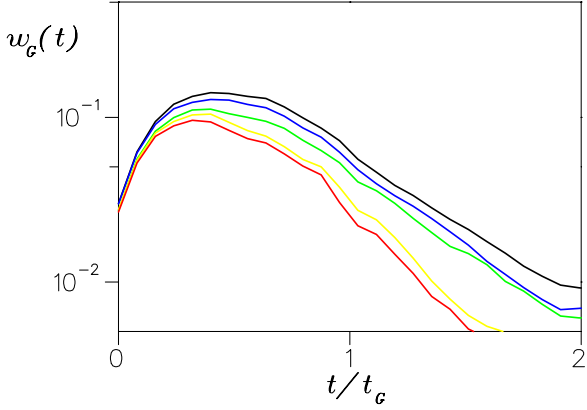}
\caption{(color online) Probability of a searched state $w_G(t)$ 
    in the Grover algorithm as a function
    of the iteration step $t$  for $n_{tot}=9$ qubits. Solid
    curves from top to bottom correspond to different numbers of units in the
    binary expansion of the number of a searched state, $n_u=1,2,4,6,8$.
    Number of quantum trajectories used in simulations is $M=2000$.
}
\label{fig2}
\end{figure}

It is interesting to note that the decay of Grover oscillations
depends on the binary representation of searched state $\vert \tau \rangle$.
Indeed, the larger is the number $n_u$ of spin-up states 
(numer of units) in the binary
representation of $\vert \tau \rangle$ the faster is the decay 
of oscillations as it is shown in Fig.~\ref{fig2}.
The physical origin of this effect is quit clear since 
in the amplitude decoherence channel
the decay $\Gamma$ takes place only for spin-up qubits.
However, even if this effect is clearly seen in 
 Fig.~\ref{fig2} in average it is not very strong since
during evolution the qubit rotates between two states
that leads to averaging of this effect. In the following
we will neglect the small deviations produced by this effect
and will analyze the averaged behaviour.

\begin{figure}
\epsfxsize=3.2in
\epsfysize=2.6in
\epsffile{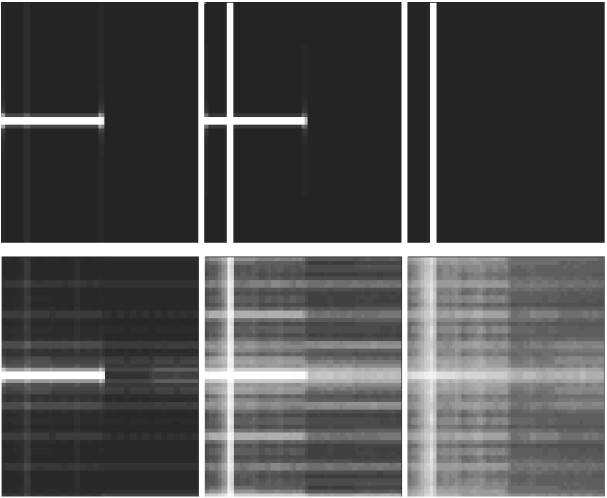}
\caption{Husimi function in the Grover algorithm at
    different number of iterations: $t=1,17,35$, from left
    to right. Top panels correspond to the ideal Grover algorithm, and bottom
    panels correspond to the single-qubit decay rate
    $\Gamma=2\cdot 10^{-4}$ ($n_{tot}=12$).
    The horizontal axis shows the computational basis
    $x=0, ...,2N-1$, while the vertical axis represents the
    conjugated momentum basis. Density is proportional to
    grayness changing from maximum (white) to zero (black).
}
\label{fig3}
\end{figure}

A pictorial image of the algorithm accuracy can be
obtained with a help of the Husimi function \cite{husimi,zhirov}
which is shown in Fig.~\ref{fig3}. In the ideal algorithm the total
probability is transfered from the initial state 
(horizontal white half line corresponding to one state of ancilla qubit)
to the searched state (vertical white line) after 35 iterations.
In presence of dissipative decoherence the probability of searched state
is significantly reduced and probability is transfered to 
the state $\vert \eta \rangle$ and 
many other states of the computational basis (Fig.~\ref{fig3}).
It is interesting to note that these transitions
give homogeneous distribution in the computational basis
(at one state of ancilla qubit) and select specific momentum
states in the momentum representation that
mainly corresponds to dissipative flips of qubits to zero state.

\begin{figure}
\epsfxsize=3.2in
\epsfysize=2.6in
\epsffile{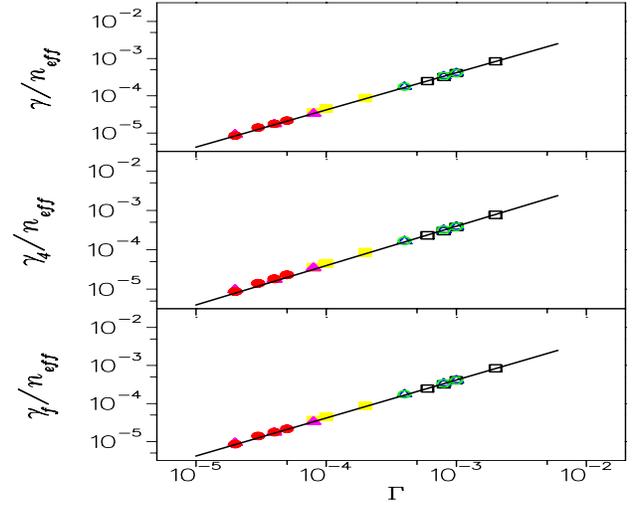}
\caption{(color online) Effective rate $\gamma$ 
   for decay of the probability of searched state
   (top), effective rate $\gamma_4$ for decay of the weight of 4-state invariant   Grover subspace (middle) and effective fidelity decay rate $\gamma_f$
   (bottom) as a function of a single qubit decay rate $\Gamma$;
   $n_{eff}=n_g n_{tot}$. Data  correspond to 
    $n_{tot}=8$ (open squares), 9 (open triangles), 10 (open circles) and
    to  $n_{tot}=12$ (full squares), 15 (full triangles), 16 (full circles).
    Full lines show the average dependence (\ref{eq6}). }
\label{fig4}
\end{figure}

\begin{figure}
\epsfxsize=3.2in
\epsfysize=2.6in
\epsffile{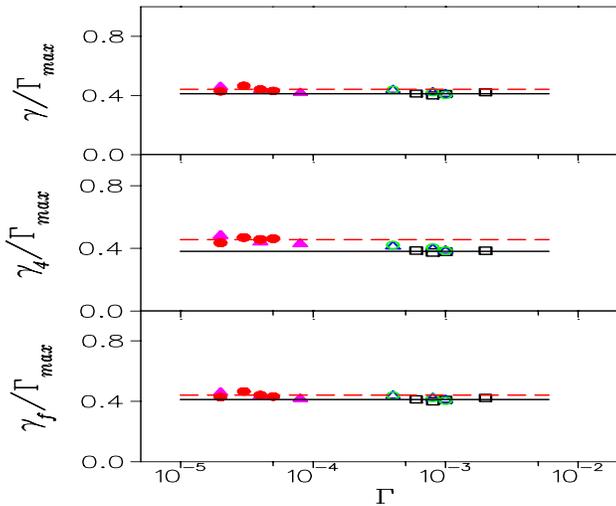}
\caption{(color online) Ratios of effective decay rates 
   $\gamma$,   $\gamma_4$ and $\gamma_f$ (as in Fig.\ref{fig4})
   to the maximum decay rate $\Gamma_{max}=\Gamma n_g n_{tot}$
   as a function of $\Gamma$.
   The dashed and full lines give the fit values of 
   constant $C$ obtained for two groups of qubits
   (dashed for  $n_{tot}=8, 9, 10$ and full
   for $n_{tot}=12, 15, 16$).
   The symbols are the same as in Fig.\ref{fig4}.
}
\label{fig5}
\end{figure}

From the data similar to those of Fig.~\ref{fig1} it is possible to
obtain the exponential decay law for $w_G, w_4, f $ ($\propto \exp(-\gamma t)$)
and find from it the corresponding decay rates $\gamma, \gamma_4, \gamma_f$.
Their dependence on the one qubit decay rate $\Gamma$
is shown in Fig.~\ref{fig4} for $n_{tot}=8, .... 16$.
In all three cases the dependence is well described by the relation
\begin{equation}
\label{eq6}
\gamma = C \Gamma n_g n_{tot} \; ,
\end{equation}
where $C \approx 0.4$ is a numerical constant.
This result gives the decay rate $\gamma$ which is close to the maximal
decay rate $\Gamma_{max}=\Gamma n_g n_{tot}$ 
which corresponds to the state with
all qubits in the up state. Quite naturally $\gamma$ is proportional
to the total number of qubits $n_{tot}$ since a flip of each
qubit from up state gives the decay of fidelity 
for the whole wave function. 
The constant $C$ is not sensitive to a variation of the number of qubits
that is illustrated in  Fig.~\ref{fig5}.
Indeed, a fit for two groups of qubits ($n_{tot}=8,9,10$
and $n_{tot}=12, 15, 16$) gives close values of $C$ (see Fig.~\ref{fig5}).
We note that $C$ is close to the value 1/2
since in average only a half of qubits is in the up state
that reduces the value of $\gamma$ by a factor 2
comparing to the maximum decay rate $\Gamma_{max}$.

The result (\ref{eq6})
is in agreement with the dependence found in \cite{carlo,lee}
for other quantum algorithms with dissipative decoherence.
This means that the decay rate relation (\ref{eq6})
gives a universal description
of dissipative decoherence in various quantum algorithms.
Therefore it is possible to compare the three classes of
quantum errors described at the beginning.
The comparison shows that the most rapid decrease of fidelity,
and thus the accuracy of quantum computation,
is produced by static imperfections.
Thus it is necessary to develop specific error correction
methods which will be able to handle effects of static imperfections
in quantum algorithms.
First steps in this direction are done recently 
in \cite{kern1,zhirovjap,kern2}.

This work was supported in part by the EC  IST-FET projects EDIQIP
and EuroSQIP and
(for OVZ) by RAS Joint scientific program "Nonlinear dynamics
and solitons".

\end{document}